\documentclass{article}
\usepackage{epsfig}
\def\beq{\begin{equation}}
\def\eeq{\end{equation}}
\def\bea{\begin{eqnarray}}
\def\eea{\end{eqnarray}}
\def\bq{\begin{quote}}
\def\eq{\end{quote}}

\def\bear{\begin{array}}
\def\ear{\end{array}}

\def\ga{\left(}
\def\dr{\right)}

\def\rar{\rightarrow}

\def\la{\langle}
\def\ra{\rangle}
\def\nin{\noindent}
\def\ba{\begin{array}}
\def\ea{\end{array}}

\def\b{\bullet}

\def\gam5{\gamma_5}

\begin{document}
\pagestyle{empty}
\sloppy
\begin{center}
\section*{ Tests of the nature and of the gluon content of the
$\sigma(0.6)$ from $D$ and $D_{s}$ semileptonic
decays}

\bigskip
{\bf Hans Guenter Dosch} \\
\vspace{0.3cm}
Institut f\"ur Theoretische Physik\\
University of Heidelberg\\
Philosophenweg\\
Germany\\
Email: h.g.dosch@thphys.uni-heidelberg.de\\

\vspace*{.5cm}
{\bf Stephan Narison} \\
\vspace{0.3cm}
Laboratoire de Physique Math\'ematique\\
Universit\'e de Montpellier II\\
Place Eug\`ene Bataillon\\
34095 - Montpellier Cedex 05, France\\
Email: qcd@lpm.univ-montp2.fr\\
\vspace*{1.5cm}

\vspace*{1.5cm}
{\bf Abstract} \\
\end{center}
We summarize the different features which show that QCD spectral sum rule
analyses of the scalar two- and three-point  functions do not
favour the $\bar uu+\bar dd$ interpretation of the broad and low mass $\sigma
(0.6)$ and  emphasize that a measurement of the
$D_s$ semileptonic decays into $\pi\pi$ can reveal in a
   model-independent way its eventual gluon component
$\sigma_B$.
The analysis also implies that one expects  an observation of the
$K\bar K$ final states from the $\sigma_B$ which may compete (if
phase space
allowed) with the one from a low mass
$\bar ss$ state assumed in the literature to be the SU(3) partner of
the observed $\sigma (0.6)$ if the latter is a
$\bar uu+\bar dd$ state.\vspace*{2mm}
\noindent
\vfill\eject
\pagestyle{plain}
\setcounter{page}{1}
\section{Introduction}
The nature of scalar mesons is an intriguing problem in QCD. Experimentally,
there are well established scalar mesons with isospin $I=1$, the
$a_0(980)$, with isospin $I=1/2$
$K^*_0(1410)$ meson,  and with isospin $I=0$, the $f_0$-mesons at
980, 1370 and 1500 MeV \cite{PDG}. Besides these
resonances there are recent experimental~\cite{Link:2001kr,KYOTO} and
theoretical~\cite{Colangelo:2001df,KYOTO} indications for a
low lying scalar isoscalar state, the famous $\sigma$.
The isoscalar scalar states are especially interesting in the
framework of QCD since, in this $U(1)_V$ channel, their interpolating
operator is the trace of the
energy-momentum tensor:
\beq
\theta_\mu^\mu\equiv \theta_g+\theta_q=\frac{1}{4}\beta(\alpha_s)
G^2+\sum_i \ga 1+\gamma_m(\alpha_s)\dr
m_i\bar\psi_i\psi_i~,
\eeq
where $G^a_{\mu\nu}$ is the gluon field strengths, $\psi_i$ is the
quark field; $\beta(\alpha_s)$ and
$\gamma_m(\alpha_s)$ are respectively the QCD $\beta$-function and
quark mass-anomalous dimension. In the chiral
limit
$m_i=0$,
$\theta_\mu^\mu$ is dominated by its gluon component $\theta_g$, like
is the case of the $\eta'$ for the
$U(1)_A$ axial-anomaly, explaining why the $\eta'$-mass does not
vanish like other Goldstone bosons for $m_i=0$. In
this sense, it is natural to expect that these $I=0$ scalar states
are glueballs/gluonia or have at least a strong
glue admixture in their wave function. QCD spectral sum sum rules
(QSSR) are an important analytical tool of
nonperturbative QCD and especially well suited to address the
question of the quark-gluon mixing since the
principal nonperturbative ingredients are the quark condensate, the
gluon condensate and the mixed quark-gluon
condensate.

\nin
In this note we summarize some essential features of previous sum rule analyses
and especially point out the relevance of semileptonic $D$ and $D_s$-decays
for obtaining information on the gluon content of the scalar mesons.
\section{Instantons and tachyonic gluon effects to the $S_2(\bar uu+\bar dd)$}
Masses and couplings of unmixed scalar $\bar qq$ mesons and
gluonia have been extensively studied in the past and more recently
\cite{SNG,SNB} reviewed using QSSR within the standard Operator
Product Expansion (OPE) of the diagonal two-point
correlator:
\beq
\psi(q^2)=i\int d^4x e^{iqx}\la 0|{\cal T}J(x)J(0)^\dagger |0\ra~,
\eeq
associated to the quark
or/and the gluonic currents.

It has been emphasized that the mass of the scalar $S_2\equiv \bar
uu+\bar dd$ meson is about 1 GeV, in agreement with
the one of the observed $a_0(980)$, and with  good $SU(2)$
symmetry implying a degeneracy between the isovector $a_0$
and isoscalar state $S_2$, while its width into $\pi\pi$ has been found to
be about 100 MeV \cite{SNB,SNG}. On the other hand, the mass of the mesons
containing a strange quark is above 1 GeV due to $SU(3)$ breaking,
which explains successfully the well-known
$\phi$--$\rho$ and
$K^*$--$\rho$ mass splittings \footnote{A more complete spectrum of 
different scalar mesons from QSSR analysis are
given in details in \cite{SNG,SNB}.}.

Contributions outside the usual OPE do not alter this result. The 
$1/q^2$-renormalon
contribution introduced by a tachyonic gluon
mass \cite{CNZ} from the
linear term of the short distance part of the QCD potential
\footnote{A linear term of the potential at all distances has been
recently proposed by 't Hooft \cite{THOOFT} as a possible way to
solve the confinement problem.} are very small. Instanton 
contributions~\cite{SHURYAK}
to the scalar current as studied in~\cite{Dosch:2002rh} improve the stability
for the sum rule of the
decay constant, but do not lower the mass of the $S_2$
{\em in a stable way}.

In Figure \ref{mass} we show the results for the mass of the 
$S_2(\bar u u + \bar d d)$
as function of the Borel/Laplace parameter $M^2$ for the leading 
contributions (a),
with the inclusion of the two loop corrections and the tachyonic mass 
contribution (b,
dashed line) and also including the instanton conribution (b, solid 
line). The continuum
threshold is 1.4 GeV$^2$ in all cases \footnote{This value is 
analogous to the ones
obtained from the QSSR analysis of some other scalar channels \cite{SNG,SNB}.}.
\setlength{\unitlength}{10mm}
\begin{figure}
\begin{center}
\epsfxsize7cm
\epsffile{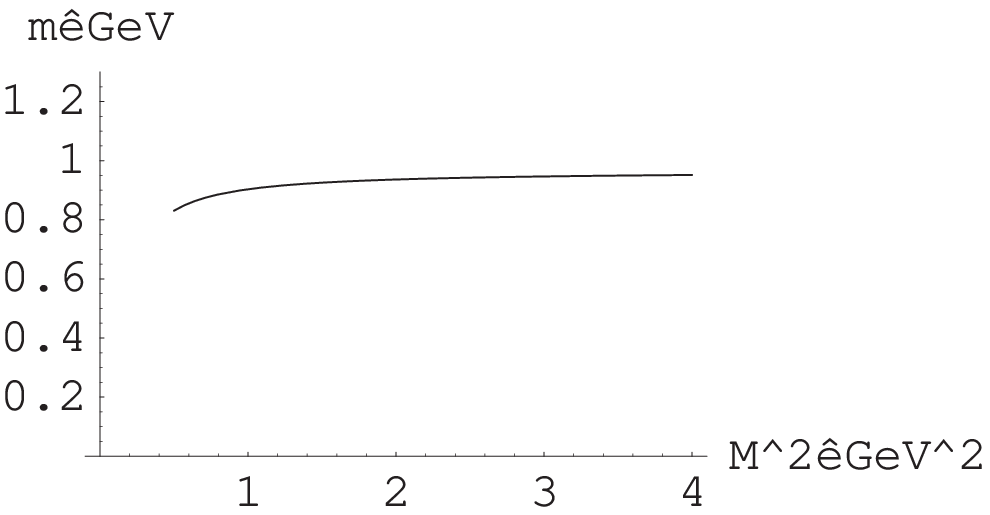}
\epsfxsize7cm
\epsffile{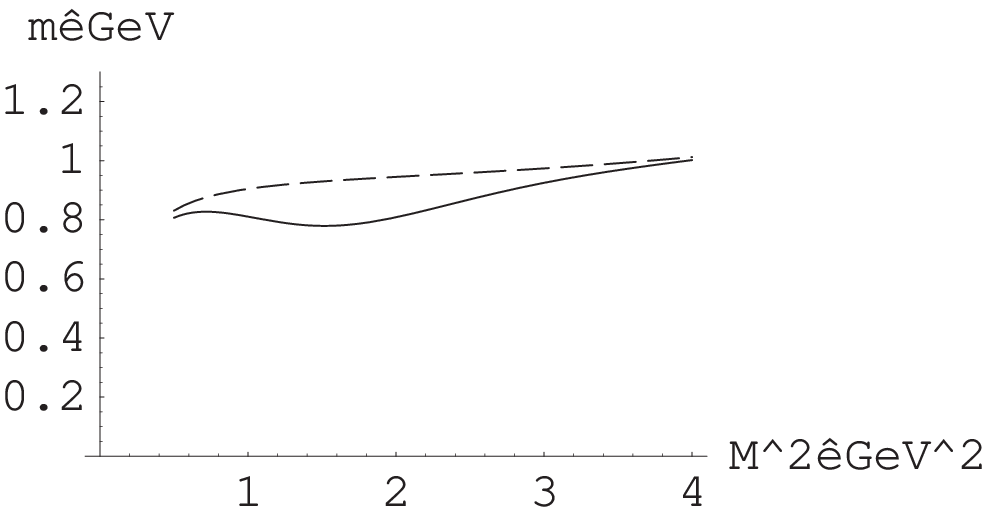}
\end{center}
\begin{picture}(0,0)
\put(4,0){a)}
\put(11,0){b)}
\end{picture}
\caption{Mass of the $S_2$ from QCDSSR. a) leading contributions. b) 
dashed: including
two loop corrections and tachyonic gluon mass; solid: plus instanton 
contribution; continuum threshold 1.4 GeV$^2$}
\label{mass}
\end{figure}
\nin
\section{Nature of the $\sigma(0.6)$ and $f_0(980)$}
 From the previous analysis, one can already conclude that {\it
the observed
$\sigma$ cannot be a pure $\bar qq$ state}.\\
$\b$ A QSSR analysis in the gluonium channel, using the
subtracted sum rule sensitive to the constant $\psi_G(0)\simeq
-16{\beta_1/
\pi}\la\alpha_s G^2\ra$, (where $\beta_1=-1/2(11-2n/3)$ and $\la
\alpha_s G^2\ra\simeq 0.07$ GeV$^4$ \cite{SNG2}) and the unsubtracted
sum
rule \cite{VENEZIA,SNB,SNG}, requires two resonances for a consistent
saturation of the two sum rules, where the lowest mass gluonium
$\sigma_B$ should be below 1 GeV. \\
$\b$ A low energy theorem for the vertex
$\la\pi|\theta^\mu_\mu|\pi\ra$ also shows that the $\sigma_B$
can be very wide with a $\pi^+\pi^-$ width of about $(0.2-0.8)$ GeV
corresponding to a mass of $(0.7-1)$ GeV and the $g_{\sigma\pi\pi}$
coupling behaves as $M^2_\sigma$. A such result shows {\it a huge
violation of the OZI rule} analogous to the one
encountered in the
$\eta'$-channel \cite{SHORE}.
\\
$\b$ From this result, a natural quarkonium-gluonium mixing (decay mixing
\footnote{This has to be contrasted with the small mass-mixing coming
from
the off-diagonal two-point function \cite{PAK}.}) scheme has been
proposed in the
$I=0$ scalar sector
\cite{BN} to explain the observed spectra and widths of the
possibly wide $\sigma (< 1$ GeV) and the narrow $f_0(0.98)$. The
data are well fitted with a nearly maximal mixing angle  $
|\theta_S|\approx 40^0~,
$
indicating that the $\sigma$ and $f_0$ have equal numbers of quark
and gluon in each of their wave functions. This mixing
scenario also implies a strong coupling of the $f_0$ to $\bar KK$
(without requiring to a four-quark $\bar ss(\bar uu+\bar
dd)$ state model) with a strength
\cite{BN}:
$
g_{f_0K^+K^-}=2g_{f_0\pi^+\pi^-}~,
$
as supported by the data. The physical on-shell $f_0$ is narrow $(<
134$ MeV) due to a destructive mixing,
whilst the $\sigma(.7\sim 1.)$ can be $(0.4\sim 0.8)$ GeV wide
(constructive mixing). Compared to the
four-quark states and/or $\bar KK$ molecules models (see e.g.
\cite{BLACK}), this quarkonium-gluonium mixing scenario includes all
QCD
dynamics based on the properties of the scale anomaly
$\theta_\mu^\mu$, which comes from QCD first principles. It is
certainly
interesting to find some further tests of this scenario, which we
propose in the following.
\section{$D_{(s)}$ semileptonic decays}
\subsection*{$S_2(\bar uu+\bar dd)$ meson productions}
A theoretically very clean way to investigate hadronic resonances is 
the analysis of
semileptonic decays of charmed mesons. Though there is much better statistics
for non-leptonic decays, the complicated final state interaction both 
on the quark
and on the hadronic level make the analysis here difficult and poses 
many puzzles
~\cite{CL02}\\
$\b$ If the scalar mesons were simple $\bar qq$ states, the
semileptonic decay width
could be calculated quite reliably using the QCD sum rule approach.
The relevant diagram is given in Fig. 2, to which  nonperturbative
contributions are added.
This has
been done with a good success for the semileptonic decays of the $D$ and $D_s$
into pseudoscalar and vector mesons \cite{SNB}. For the production of
a pseudoscalar or
scalar $\bar qq$ states several groups \cite{Dosch:2002rh},
\cite{Ball:1991bs} predict all form factors to be:
$
f_+(0)\approx 0.5~,
$
where a similar value has been obtained from a completely independent
approach~\cite{Gatto:2000hj} based on the
constituent quark model. This
yields a decay rate:
\beq
\Gamma(D\rar S_2 l\nu)=(8\pm 3)10^{-16} ~{\rm GeV}~,
\eeq
for $M_{S_2}\simeq 600$ MeV.
\begin{figure}[hbt]
\begin{center}
\epsfxsize4cm
\epsffile{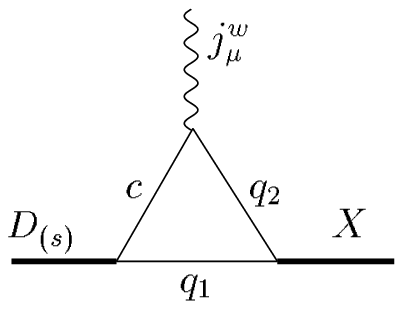}
\end{center}
\caption{Schematic picture of the semileptonic vertex}
\end{figure}
\\
\nin
$\b$ However, because of the enigmatic nature of the  $\sigma$,  one
has also considered,  in ~\cite{Dosch:2002rh}, the
case that the quark-antiquark current occuring in Fig. 2 does not couple to
a resonance but rather to an uncorrelated quark-antiquark pair. In that case
the decay rate is reduced by a factor 2, but in the spectral distribution
nevertheless there is a broad bump visible with a maximum near the presumed
$\sigma$ mass of 600 MeV (see Figure 4 of ~\cite{Dosch:2002rh}).
Unfortunately even in high stastistics
experiments the estimated decay rates of the $D$-meson are at the
edge of observation since the
decays into an isocalar are CKM-suppressed due to the $c$-$q$ transition at
the weak vertex.
\subsection*{Scalar gluonium and/or $\bar ss$ productions}
The
diagrams for a semileptonic decay into a gluonium state,
are given in Fig. 3. Unfortunately the evaluation of these diagrams
is more involved than in the $\bar qq$ case.
Therefore, we can give only semi-qualitative results which however are
model independent.\\
$\b$ The only way to obtain a non-CKM suppressed isoscalar is to look at the
semileptonic decay of the $D_s$-meson. As shown in Figs. 2 and 3,  the
quark $q_1$ is a strange one and an isoscalar $s \bar s$ or gluonium
state can be
formed. \\
\begin{figure}[hbt]
\begin{center}
\epsfxsize8cm
\epsffile{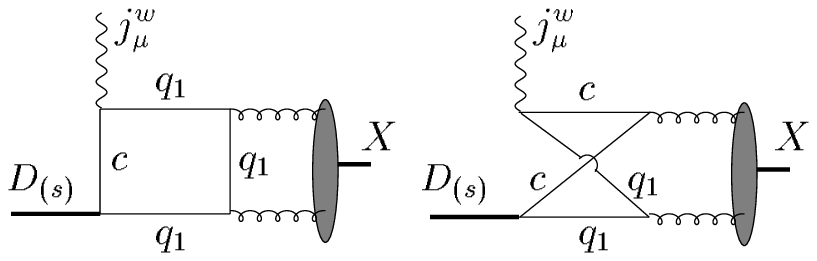}
\end{center}
\caption{Glue ball formation in semileptonic decays}
\end{figure}\\
\nin
$\b$ If the $\bar ss$ is relatively light ($<1$ GeV), which might be
the natural partner of the $\bar
uu+\bar dd$ interpreted  to be a $\sigma (0.6)$ as often used in the
literature, then, one should produce a
$\bar KK$ pair through the isoscalar $\bar ss$ state. The QSSR
prediction for this process is under quite good
control \cite{SNB,Dosch:2002rh}. The
non-observation of this process will disfavour the
$\bar qq$ interpretation of the $\sigma$ meson. \\
$\b$ If a gluonium state is formed it will decay with even strength
into $\pi\pi$ and a $K\bar K$ pairs. Therefore a gluonium
formation in semileptonic $D_s$ decays should result in the decay patterns:
\beq
D_s\rar \sigma_B\ell \nu\rar \pi \pi \ell \nu ~~~~~~~~~~~~~~
D_s\rar\sigma_B\ell \nu\to K \bar K  \ell \nu~,
\eeq
with about the same rate up to phase space factors.
{\it The observation of the semileptonic $\pi\pi$ decay of the $D_s$
would be a unique
sign for glueball formation.}\\
$\b$ A semi-qualitative estimate of the above rates can be obtained
by working in the large heavy quark mass limit $M_c$. Using
e.g. the result in ~\cite{Dosch:2002rh}, the one for light $\bar qq$
quarkonium production behaves as:
\beq
\Gamma(D_s\rar S_q(\bar qq)~l\nu)\sim |V_{cq}|^2G^2_F M^5_c |f_+(0)|^2~.
\eeq
$\b$ For the $\sigma_B(gg)$ production, we study the $1/M_c$
behaviour of the $WW gg$ box diagram in Fig. 3,
where it is easy to find that the dominant (in $1/M_c$) contribution
comes from the one in Fig. 3a. Therefore the production
amplitude can be described by the Euler-Heisenberg effective interaction :
\beq
{\cal L}_{eff}\sim \frac{g_W\alpha_s}{p^2M_c^2}F_{\mu\nu}
F^{\mu\nu}G_{\alpha\beta}G^{\alpha\beta}+ {\rm
permutations}+{\cal O}(\frac{1}{M_c^4})
\eeq
where $g_W$ is the electroweak coupling and $p^2\simeq
M^2_{\sigma_B}$ is the typical virtual low scale entering into the box
diagram. Using dispersion techniques similar to the one used for
$J/\Psi\rar \sigma_B \gamma$ processses \cite{NSVZ,VENEZIA,SNB}, one
obtains, assuming a $D_s$ and
$\sigma_B$-dominances:
\beq
\Gamma(D_s\rar\sigma_B(gg)~l\nu)\sim |V_{cs}|^2G^2_F M^3_c
\frac{1}{M^4_cM_\sigma^4} |\la 0|\alpha_s G^2|\sigma_B\ra|^2
\eeq
The matrix element $\la 0|\alpha_s G^2|\sigma_B\ra $ is by definition
proportional to $f_\sigma M^2_\sigma$, where $f_\sigma$ is
hopefully known from two-point function QSSR analysis
\cite{VENEZIA,SNG,SNB}. Using $f_\sigma\approx 0.8$ GeV, one
then deduce:
\beq
{\Gamma(D_s\rar\sigma_B(gg)~l\nu)\over \Gamma(D_s\rar S_q(\bar
qq)~l\nu)}\sim {1\over |f_+(0)|^2}\ga{f_\sigma \over M_c}\dr^2\sim
{\cal O}(1)
\eeq
This qualitative result indicates that {\it the gluonium production
rate can be of the same order as the $\bar qq$ one} contrary to
the na\"\i ve perturbative expectation of the $\alpha_s^2$
suppression rate. This result being {\it a consequence of the OZI-rule
violation of the $\sigma_B$ decay.}\\
However, it also indicates that, due to the (almost) universal
coupling of the $\sigma_B$ to Goldstone boson pairs, one also expects
a production of the $K\bar K$ pairs, which can compete with the one
from $\bar ss$ quarkonium state, and again renders more
difficult the identification of the such state $\bar ss$ if allowed 
by phase space.
\section{Conclusions}
After reminding the different features from QCD spectral analysis of
the scalar two- and three-point functions which do not
favour the $\bar qq$ interpretation of the broad and low mass $\sigma
(0.6)$, we have emphasized that a measurement of the
$D_s$ semileptonic decays into $\pi\pi$ can reveal in a clean, unique
and model-independent way the eventual gluon component of
the
$\sigma$ meson. The analysis also implies that one expects  an
observation of the $K\bar K$ final states from the $\sigma_B$
which can compete (if any) with the one expected from a
$\bar ss$ state assumed in the literature to be the SU(3) partner of
the observed $\sigma(0.6)$ often interpreted as a
$\bar uu+\bar dd$ state.
\section*{Acknowledgements} S.N. wishes to thank the Alexander Von
Humboldt Foundation which provides the grant
for visiting the Institut f\"ur Theoretische Physik of the University
of Heidelberg where this work has been initiated.

\end{document}